\documentclass[shortnote,twocolumn]{jpsj3}
\usepackage{txfonts}
\usepackage{float}
\usepackage{graphicx}

\title{Academic Meeting Scheduling Using an Antiferromagnetic Potts Model}

\author{Kazue Kudo\thanks{kudo@is.ocha.ac.jp}}
\inst{Department of Information Sciences, Ochanomizu University, Bunkyo-ku, Tokyo 112-8610, Japan} 

\abst{Scheduling parallel sessions of an academic meeting is a complicated task. 
If each presentation is assigned to an appropriate session, an
antiferromagnetic Potts model can be used for semi-automatic timetabling.
The timetabling method proposed here is based on graph coloring and includes additional constraints to be considered in a practical situation.
We examine the feasibility of semi-automatic timetabling in some practical examples.}

\begin{document}
\maketitle

There are often many parallel sessions in an academic meeting, and scheduling them is a complicated task.
After collecting the information about all presentations,
organizers need to (1) organize presentations into appropriate sessions;
(2) assign a timeslot to each session; and
(3) allocate a meeting room to each session.
The most complicated process is probably assigning timeslots
as some constraints need to be considered.
For example, presentations by the same presenter in different sessions should not be in the same timeslot.
Further, sessions with the same topic and those with largely overlapped audiences should not be in the same timeslot.
Timetabling, i.e.,  assigning timeslots, under such complicated constraints is a practical application of graph coloring,
which is an NP-complete combinatorial problem~\cite{Garey79}.

Graph coloring, especially the coloring of random graphs,
has attracted considerable interest from different fields such as 
mathematics, computer science, and statistical physics\cite{Wu82}.
Given a graph (or network) and $q$ available colors,
the problem consists in coloring the nodes (or vertices) such that
no pair of nodes connected by an edge (or link) has the same color.
In the timetabling problem,
node, color, and edge correspond to session, timeslot, and constraint, respectively.
For large random graphs,
there exists a critical average connectivity (vertex degree) beyond which a
graph becomes uncolorable with $q$ colors.
The colorable/uncolorable (COL/UNCOL) transition and some other phase transitions were extensively studied by using antiferromagnetic $q$-state Potts models~\cite{mulet02,vanMourik02,braunstein03,krzakala04,zdeborova07,krzakala07,krzakala08}.
In the antiferromagnetic Potts model, energy contribution arises
when interacting (neighboring) nodes are in the same state,
while no contribution arises from neighboring nodes in different states.
Thus, finding the ground state of the Potts model is equivalent to graph coloring.

In this study, we focus on timetabling (assigning timeslots to sessions), which is a part of academic meeting scheduling.
Here, we assume that all presentations are already organized into appropriate sessions.
Two additional constraints are considered other than the constraint between sessions that should avoid the same timeslot.
One is avoidance constraint on timeslots:
some sessions often need to avoid particular timeslots for some reason.
The other is the number of meeting rooms:
the number of available rooms is often limited.
These constraints are incorporated into an antiferromagnetic Potts model with additional terms.
The model enables semi-automatic timetabling in the sense that
the task of assigning timeslots no longer bother organizers.
The aim of this article is to propose a key idea to realize semi-automatic timetabling.

Suppose that variable $s_i$ is the timeslot assigned to session $i$, the number of available timeslots is $q$, and the number of sessions is $N$
($s_i=1,\ldots, q$, $i=1,\ldots, N$).
The Hamiltonian to be considered here reads
\begin{equation}
 H = \sum_{i,j}J_{ij}\delta(s_i,s_j)
  + \sum_{i=1}^N\sum_{k=1}^q w_{ik}\delta(s_i,k)
  + \sum_{k=1}^q \max(N_k-R_k,0),
\label{eq:H}
\end{equation}
where $\delta(s_i,s_j)$ denotes the Kronecker delta.
The first term of the right-hand side (RHS) in Eq.~\eqref{eq:H} represents the constraint between sessions that should avoid the same timeslot.
If no constraint exists between sessions $i$ and $j$, $J_{ij}=0$;
otherwise, $J_{ij}>0$.
The second term of the RHS in Eq.~\eqref{eq:H} represents the avoidance constraint.
If session $i$ needs to avoid timeslot $k$, $w_{ik}>0$; otherwise $w_{ik}=0$.
The third term of the RHS in Eq.~\eqref{eq:H} comes from the limited number of meeting rooms.
$N_k=\sum_{i=1}^N\delta(s_i,k)$ is the number of sessions in timeslot $k$,
and $R_k$ is the number of available rooms in timeslot $k$.
The third term gives a contribution only when $N_k>R_k$.

\begin{figure}[tb]
 \begin{center}
  \includegraphics[width=8cm]{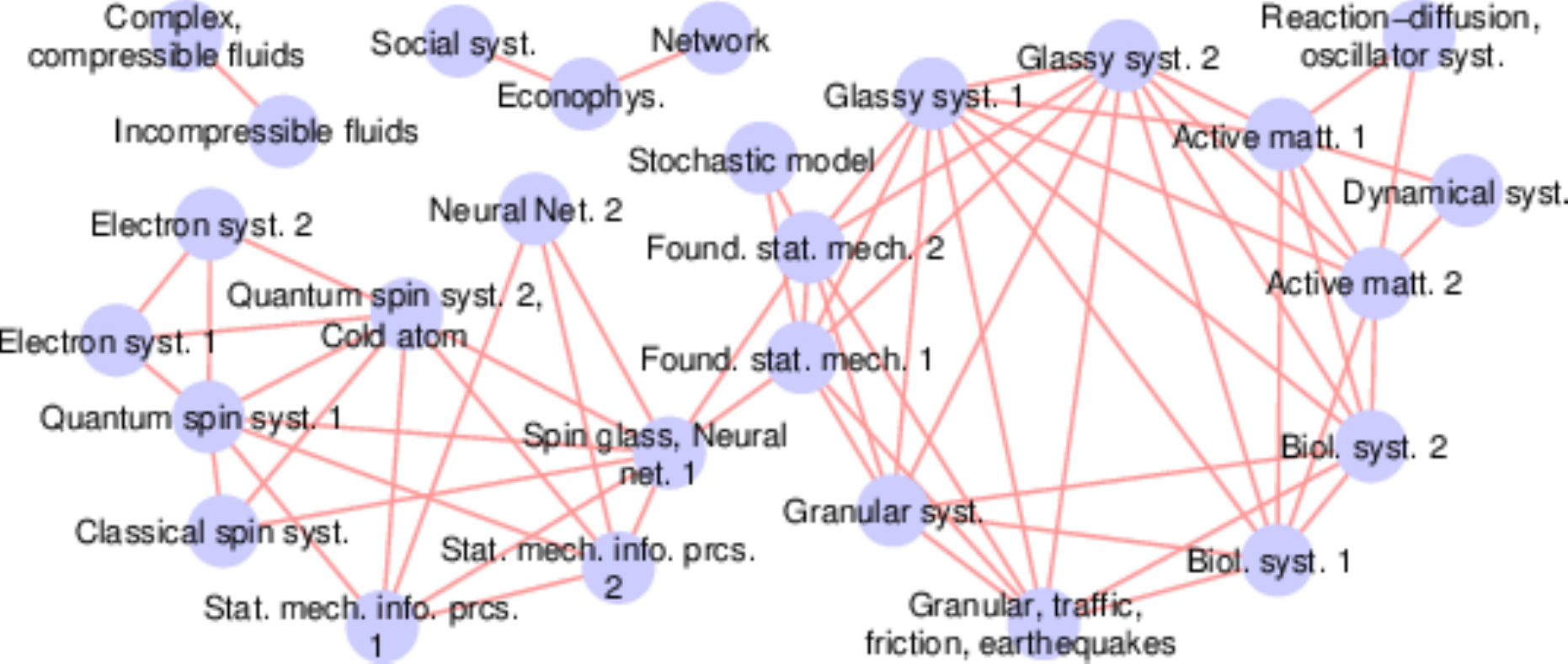}
 \end{center}
 \caption{(Color online) Graph of constraints between oral sessions (excluding special sessions such as symposia) in Division 11 at 2017 Annual Meeting of the JPS.
 Sessions with no edges are not shown.}
 \label{fig:graph}
\end{figure}

Now, we examine practical examples: meetings of the Physical Society of Japan (JPS). 
In particular, we focus on sessions of Division 11 (Fundamental Theory of Condensed Matter Physics, Statistical Mechanics, Fluid Dynamics, Applied Mathematics, Socio- and Econophysics),
which is one of the largest divisions in the JPS.
At each Annual/Autumn (or Spring) Meeting, over 200 oral presentations are given in Division 11 in four days.
Presentations are organized into sessions, based on the primary keyword of each presentation.
Division 11 has some rules about constraints between sessions.
The graph in Fig.~\ref{fig:graph} illustrates the rules (as of March 2017).
Organizers have to assign a timeslot to each session so that the constrains are satisfied.
The number of available timeslots is $7$ for an Annual Meeting and $8$ for an Autumn (or Spring) Meeting.

\begin{table}[tb]
 \caption{Properties of graphs of constraints between oral sessions (excluding special sessions) in Division 11 of the JPS. The number in parentheses is the number of oral sessions including ones with no edges. Overlaps of presenters are not taken into consideration.}
 \label{t1}
 \begin{center}
  \begin{tabular}{ccccc}
   \hline
   & 2017   & 2016   & 2016   & 2015 \\
   & Annual & Autumn & Annual & Autumn \\
   \hline
   The number of nodes & 27 (31) & 26 (31) & 28 (31) & 24 (29) \\
   Average connectivity & 4.815 & 4.846 & 3.929 & 4.167 \\ 
   Maximum clique & 6 & 6 & 5 & 6\\ 
   \hline
  \end{tabular}
 \end{center}
\end{table}

The graph of constraints between sessions at each meeting has similar properties to each other.
Table~\ref{t1} shows properties of the graphs of the last four major meetings.
There are about 30 oral sessions (excluding special sessions such as symposia) in Division 11.
Some of these sessions have no constraint between other sessions. 
The average connectivity is less than 5, which is far lower than the coloring threshold $c_s$ for random graphs.
For instance, in the $q$-coloring problem on Erd\"os-R\'enyi graphs,
$c_s\simeq 24.255$ for $q=7$ and $c_s\simeq 30.355$ for $q=8$~\cite{zdeborova07}.
The number of nodes in the maximum clique is less than 7 or 8.
Here, a clique is the subgraph that is a complete graph.
For coloring a graph with a $k$-node clique, at least $k$ colors are needed.
These facts indicate that a typical graph of constraints such as that listed in Table~\ref{t1} is colorable unless additional constraints are considered.


We investigate the effects of the additional constraints,
namely, avoidance constraints and the limited number of meeting rooms,
using the Potts model given by Eq.~\eqref{eq:H}.
To simplify the problem, the values of $J_{ij}$ and $w_{ik}$ are restricted to $0$ or $1$.
In this situation, $H=0$ if and only if all constraints are satisfied.
Although $J_{ij}$ is given from the graph of constraints between sessions,
$w_{ik}$ is given randomly.
In other words, each session is supposed to avoid each timeslot with a certain probability.
For given parameters $J_{ij}$, $w_{ik}$, and $R_k$, we look for a combination of variables $s_i$ that gives $H=0$, using a simulated annealing method.

\begin{figure}[tb]
 \begin{center}
  \includegraphics[width=8cm]{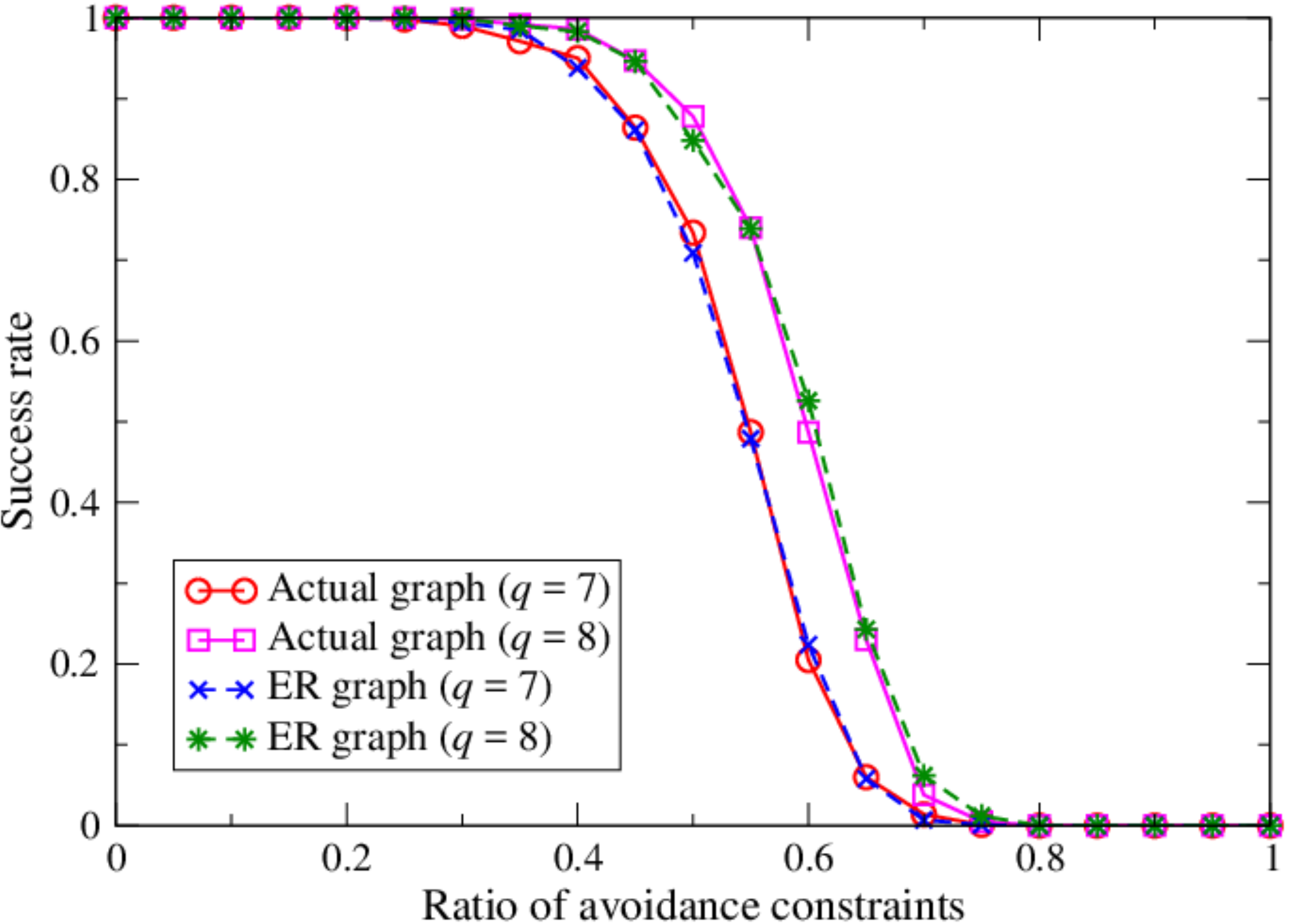}
 \end{center}
 \caption{(Color online) Success rates of timetabling plotted as functions of the ratio of avoidance constraints.
 The number of nodes is $N=31$, and the number of available timeslots is $q$.
 The total number of meeting rooms is given by $N$.
 Actual graph is the graph, which is shown in Fig.~\ref{fig:graph}, of constraints between oral sessions in Division 11 at 2017 Annual Meeting of the JPS.
 It includes nodes with no edges.
 ER graph is an Erd\"os-R\'enyi graph with the same number of nodes and average connectivity as the actual graph.}
 \label{fig:w}
\end{figure}

Avoidance constraints strongly influence the coloring of graphs,
which is illustrated in Fig.~\ref{fig:w}.
The ratio of avoidance constraints is the ratio that $w_{ik}=1$.
The success rate is the ratio of the samples that achieved $H=0$ among all 1000 samples with different combinations of $w_{ik}$'s (and also different combinations of edges in Erd\"os-R\'enyi graphs).
The graph shown in Fig.~\ref{fig:graph} is used as an actual graph.
Here, however, it includes nodes without any edges.
Even when the ratio of avoidance constraints is $0.5$, the success rates are above $0.5$, although those of $q=7$ are lower than those of $q=8$, which is as expected.
Almost the same results are obtained when we use Erd\"os-R\'enyi graphs with the same number of nodes and average connectivity as the actual graph.
In contrast to avoidance constraints, dependence on the number of meeting rooms was negligible, which is not demonstrated here.

The results in Fig.~\ref{fig:w} imply that the antiferromagnetic Potts model given by Eq.~\eqref{eq:H} leads to successful scheduling unless constraints are severe.
In some cases, however, not all constraints are satisfied.
For example, if the size of the maximum clique is larger than the number of available timeslots, some constraints are certainly unsatisfied.
It is also difficult to find a solution with $H=0$ when the ratio of avoidance constraints is high.
If constrains are severe, it is sometimes good enough to find a solution with the minimum $H$ in a practical situation. 

In conclusion, the antiferromagnetic Potts model was proposed to realize semi-automatic timetabling in an academic meeting.
Simple numerical simulations demonstrated that the model is helpful for timetabling in typical cases where the number of sessions and the ratio of avoidance constraints are similar to those of Division 11 of the JPS~\cite{memo}.
Although the small-scale simulations are useful for practical applications, 
they are insufficient to generalize the problem.
To discuss the effectiveness of the model further, more powerful methods such as the cavity method~\cite{Mezard2001,Mezard2003}, exchange Monte Carlo~\cite{Hukushima96}, and simulated tempering~\cite{Marinari92,Sakai16} should be used.
We hope that this article will stimulate the statistical physics community and that automatic scheduling of a whole academic meeting will be realized in the future.

\begin{acknowledgment}


The author would like to thank Program Committee members of Division 11 of the Physical Society of Japan for stimulating discussion.
\end{acknowledgment}


\end{document}